\documentclass[preprint,12pt,3p]{elsarticle}
\usepackage{amssymb}
\usepackage{amsmath,amsfonts}
\usepackage{hyperref}
\usepackage{natbib}
\usepackage{tikz}
\usepackage{algorithm,algorithmic}
\usepackage{booktabs}
\usepackage{multirow}

\newcommand{\solidqed}{\hfill $\blacksquare$}
\newtheorem{theo}{\bf Theorem}
\newtheorem{lem}[theo]{\bf Lemma}
\newtheorem{coro}[theo]{\bf Corollary}

\newtheorem{obse}[theo]{\bf Observation}

\begin{document}

\begin{frontmatter}



\title{Prediction of Multiscale Features Using Deep Learning-based Preconditioner-Solver Architecture for Darcy Equation in High-Contrast Media\footnotetext{This work was funded by the Postgraduate Research Scholarship of Xi'an Jiaotong-Liverpool University (FOSA2412003). The research of Zhengkang He is supported by the Natural Science Foundation of Shandong Province (ZR2023QA069) and the Natural Science Foundation of China (12301528).}} 


\author[xjtlu]{Jie Chen\corref{corresponding}}
\ead{Jie.Chen01@xjtlu.edu.cn}
\author[xjtlu,uol]{Peiqi Li}
\ead{Peiqi.Li23@student.xjtlu.edu.cn}
\author[ytu]{Zhengkang He}
\ead{zkhe@ytu.edu.cn}
\author[uol]{Simon Hands}
\ead{Simon.Hands@liverpool.ac.uk}

\cortext[corresponding]{Corresponding author.}
\affiliation[xjtlu]{organization={School of Mathematics and Physics, Xi'an Jiaotong-Liverpool University},
            postcode={215123},
            city={Suzhou},
            country={China}}
\affiliation[uol]{organization={Department of Mathematical Sciences, University of Liverpool},
            postcode={L69 3BX},
            city={Liverpool},
            country={UK}}
\affiliation[ytu]{organization={School of Mathematics and Information Science, Yantai University},
            postcode={264005},
            city={Yantai},
            country={China}}

\begin{abstract}
Modeling subsurface fluid flow in heterogeneous porous media is challenging due to the multiscale complexity of permeability fields. Traditional methods enable multiscale modeling but suffer from high computational costs and limited scalability. To address these challenges, we propose Fourier Preconditioner-based Hierachical Multiscale Network (FP-HMsNet), a deep learning-based hierarchical preconditioner-solver framework that integrates a spectral preconditioner with a multiscale neural solver to efficiently predict multiscale basis functions. Our approach leverages a Fourier-based preconditioner to transform permeability fields, enhancing global and local feature learning while reducing computational complexity. A multiscale convolutional network further refines feature extraction, improving accuracy. Theoretical analysis confirms the stability, error bounds, and convergence of FP-HMsNet under high-contrast conditions. Numerical experiments demonstrate its superior accuracy and efficiency compared to numerical and existed deep learning-based methods. Ablation studies highlight the importance of spectral preconditioning and multiscale feature extraction. By significantly reducing computational costs, FP-HMsNet enables real-time decision-making in oil and gas reservoir management, groundwater contamination assessment, and subsurface energy storage, making it a valuable tool for geoscience and engineering applications.
\end{abstract}



\end{frontmatter}

\section{Introduction}
Modeling of porous media for subsurface fluid flow holds critical importance in hydrocarbon resource exploration and development, as well as in subsurface seepage prediction. Natural porous media systems frequently exhibit significant heterogeneity, requiring explicit representation of their complex fracture networks during geological modeling. These fracture systems demonstrate substantial variations in both spatial distribution and geometric characteristics, resulting in physical models that encompass multiple spatial scales. Within such multiscale systems, conventional large-scale modeling approaches often experience compromised accuracy due to the presence of subscale fractures, while exclusive reliance on fine-scale discretization becomes prohibitively computationally demanding. This inherent multiscale challenge necessitates the implementation of specialized multiscale techniques for effective simulation of fluid transport phenomena in heterogeneous porous formations.

Multiscale techniques find extensive applications in modeling subsurface fluid flow through heterogeneous porous media. Classical methodologies, including the conservative multiscale finite volume method (MsFVM) \citep{hadi2008Iterative,jenny2005Adaptive}, the mortar multiscale method \citep{ganis2017A}, the multiscale finite element method (MsFEM)\citep{chen2003A} with its variants, mixed multiscale finite element method \citep{aarnes2004use,chen2003mixed}, and generalized multiscale finite element method (GMsFEM) \citep{chung2015mixed,efendiev2013generalized} have been proposed. These methods effectively bridge microscopic structural features - such as fracture networks and pore-scale geometries — with macroscopic flow predictions through systematic homogenization procedures. \citep{chung2015residual} extended this framework by integrating adaptive online basis enrichment, enabling efficient resolution of nonlinear, time-dependent flow phenomena in complex geological formations.

Homogenization methods, which transform complex media with multiscale heterogeneity into equivalent homogeneous media, are widely applied in simulation of subsurface fluid flow \citep{azad2021hierarchical}. The fundamental assumption of these methods is that, through appropriate mathematical treatment, a complex multiscale model can be approximated by an equivalent homogeneous model at the coarse-grid scale without significantly losing simulation accuracy. In the homogenization process, statistical means, coefficients of variation, or more advanced homogenization techniques such as GMsFEM, are typically employed to capture and simplify the multiscale characteristics of the media. We assume that the permeability field in our study is a homogenized two-dimensional matrix, realized through multiscale analysis of the field.

In complex porous media conditions, preconditioning techniques are widely used to improve the efficiency of numerical methods. Especially, when dealing with multiscale problems with high-contrast coefficients, preconditioner can significantly accelerate the convergence and reduce the computational overhead, and become an effective tool to solve partial differential equations (PDEs). \cite{efendiev2013generalized} proposed an adaptive preconditioning method to construct rough spaces by selecting eigenvectors from local eigenvalue problems, which showed significant advantages in the solution of Darcy flow in porous media. \cite{fu2024efficient} used a two-grid preconditioner based on generalized multiscale space, which mainly improves the efficiency and robustness of solving Darcy problem by constructing non-standard coarse spaces containing local heterogeneity information. The preconditioner uses the feature function generated by local spectral problems to enrich the rough space and smooth it with ILU decomposition, which significantly improves the convergence rate in high-contrast porous media. The utilization of preconditioners makes PDE solvers more efficient and stable in different problem situations \cite{liu2024learning,yang2019two,fu2024edge,ye2024robust,vasilyeva2024adaptive,ye2025highly}.

In recent years, deep learning has made significant progress in solving PDEs, involving from purely data-driven approaches to hybrid methods that integrate physical constraints to enhance computational efficiency and generalization capabilities. Various neural network architectures have been introduced into PDE solvers, including convolutional neural networks (CNNs) \cite{choubineh2022innovative,choubineh2023deep}, Transformer \cite{vaswani2017attention,geng2024swin}, and Fourier neural operators (FNOs) \cite{kovachki2021universal,li2023fourier,li2020fourier}, with FNOs demonstrating strong generalization capabilities in complex physical simulations. To address the challenge of capturing high-frequency information in multiscale problems, \cite{liu2024mitigating} proposed the hierarchical attention neural operator (HANO), which mitigates spectral bias through a multilevel self-attention mechanism and achieves high-accuracy derivative approximation in high-contrast media flow. However, the high computational cost of this method limits its applicability to large-scale problems. To improve computational efficiency, \cite{spiridonov2025prediction} integrated the online GMsFEM with neural networks to rapidly predict local multiscale basis functions, significantly reducing computational overhead in nonlinear flow simulations. These studies indicate that future advancements in PDE solvers will focus on the deep integration of physics-informed constraints and data-driven approaches, the balanced optimization of computational efficiency and accuracy, and the synergistic learning of high- and low-frequency features, paving the way for broader applications of deep learning in scientific computing.

Accurately modeling subsurface fluid flow in highly heterogeneous porous media remains hindered by the inherent multiscale complexity of natural permeability fields. Conventional numerical approaches, such as MsFEM, often face a critical trade-off: while capable of resolving fine-scale heterogeneities via basis functions, they suffer from prohibitive computational costs and limited scalability, particularly in high-contrast or fractured systems. These challenges are exacerbated by the need to balance global flow pattern predictions with localized heterogeneities — a gap that persists despite advances in data-driven methods. Such limitations motivate the development of hybrid frameworks that synergize deep learning with multiscale physics to enhance both efficiency and accuracy.

In this study, we propose Fourier Preconditioner-based Hierachical Multiscale Network (FP-HMsNet), a ‌preconditioner-solver framework that jointly optimizes computational efficiency and multiscale fidelity through a Fourier-enhanced architecture. By embedding a data-driven spectral prior directly into the network, our model achieves ‌linear time complexity‌ $\mathcal{O}(N^2\log N)$ and ‌sublinear memory scaling‌ with domain size $N^2$, a critical advancement for edge-deployed subsurface simulations. Rigorous analysis demonstrates that FP-HMsNet significantly reduces approximation error through its hierarchical Fourier-convolution fusion mechanism, while theoretical guarantees on stability (via Lipschitz-constrained layers) and convergence (exponentially decaying residuals) are provided in the variational formulation. The model can maintain sublinear computational efficiency under computational power constraints, and this lightweight property allows it to be directly deployed in low-power edge computing units such as downhole sensors and embedded geological monitoring devices, significantly improving the engineering feasibility of real-time pore pressure prediction and delay-sensitive industrial systems.

The remainder of this study provides a detailed exposition of our study. Section \ref{preliminaries} introduces the background of our problem, including the fundamental theory of Darcy’s equation and the construction of multiscale basis functions based on the mixed GMsFEM. Section \ref{method} presents a comprehensive description of our proposed method, covering the dataset generation and construction process, as well as the theoretical foundations related to our model. Section \ref{theorem} evaluates the effectiveness of our model in terms of computational complexity, error analysis, stability, and convergence. Section \ref{results} analyzes the numerical experiment results and includes an ablation study to demonstrate the effectiveness of the proposed preconditioning method. Section \ref{discussion} summarizes the findings and discusses their implications, while Section \ref{conclusion} concludes the study and outlines potential directions for future research.

\section{Preliminaries}
\label{preliminaries}

\subsection{Model Problem}
Consider the following Darcy flow problem for the unknown pressure field $p$ in a bounded domain $\Omega\subset\mathbb{R}^2$:
\begin{equation}
	\begin{cases}
		-\text{div}(\kappa \nabla p)&=f~~\text{in}~\Omega, \\
		\kappa \nabla p \cdot \mathbf{n}&=0~~\text{on}~\partial\Omega,
	\end{cases}
	\label{darcy_2nd_order}
\end{equation}
where $\partial\Omega$ is the Lipschitz continuous boundary, $\mathbf{n}$ refers to the outward unit normal vector on $\partial\Omega$, $\kappa$ is the high-contrast permeability field with highly heterogeneous features, $f$ is the source term that satisfies the compatibility condition $\int_\Omega f~dx=0$, and there exists restriction $\int_\Omega p~dx=0$ to ensure the uniqueness of the solution. 

Introduce the velocity variable:
\begin{equation}
	\mathbf{u}=-\kappa \nabla p, \notag
\end{equation}
the model (\ref{darcy_2nd_order}) can be transformed into a 1st order form:
\begin{equation}
	\begin{cases}
		\kappa^{-1}\mathbf{u}+\nabla p &= 0~\text{in}~\Omega,~~~~(\text{Darcy's Law})\\
		\text{div}(\mathbf{u})&= f~\text{in}~\Omega,~~~~(\text{Mass Conservation})\\
		\mathbf{u}\cdot\mathbf{n} &= 0~\text{on}~\partial\Omega.~~(\text{No-Flux Boundary Condition})
	\end{cases}
	\label{darcy_1st_order}
\end{equation}

Let $\mathcal{T}^h$ denote a conforming and shape-regular fine-grid partition of $\Omega$ into quadrilaterals with size $h$, and $\mathcal{T}^H$ denote a quasi-uniform coarse-grid partition of $\Omega$, such that each coarse element is a connected collection of fine elements with size $H$. 
Figure.\ref{fig_mesh example} illustrates an example of the two-scale grid, where each coarse element contains $3\times 3$ fine elements.
\begin{figure}[h]
	\centering
	\begin{tikzpicture}[scale=0.4] 
		\draw[gray!50] (0,0) grid (9,9);
		
		\draw[very thick] (3,0) -- (3,9); 
		\draw[very thick] (6,0) -- (6,9);
		\draw[very thick] (0,3) -- (9,3);
		\draw[very thick] (0,6) -- (9,6);
		
		\fill[blue!20] (0,0) rectangle (3,3);
		
		\draw[gray!70, thick] (0,0) grid (3,3);
		
		\node at (1.5,1.5) {\Huge$T$};
		
		\fill[orange!60] (7,8) rectangle (8,9);
		\draw[dashed, thick] (7,8) rectangle (8,9);
		\node at (7.5,8.5) {$t$};
	\end{tikzpicture}
	\caption{An example of the two-scale grid. $t$ is a fine element in $\mathcal{T}^h$, and $T$ is a coarse element in $\mathcal{T}^H$.}
	\label{fig_mesh example}
\end{figure}

Denote $\mathcal{E}_H:=\bigcup_{i=1}^{N_e} E_i$ be the set of all edges of $\mathcal{T}^H$, where $N_e$ refers to the number of coarse edges.

Let 
\begin{equation}
	\text{H}(\text{div},\Omega)=\{\mathbf{v}\in (L^2(\Omega))^2:\nabla \cdot \mathbf{v} \in L^2(\Omega) \},
\end{equation}
and define
\begin{equation}
	V=\text{H}(\text{div},\Omega),\ \ \ W=L^2(\Omega). \notag
\end{equation}

For the fine-grid solution of (\ref{darcy_1st_order}) on $\mathcal{T}^h$, we use the lowest-order Raviart-Thomas ($\text{RT}_0$) mixed finite element spaces, given by
\begin{align}
	V_h &= \{\mathbf{v}_h \in V: \mathbf{v}_h|_t=(b_tx_1+a_t, d_tx_2+c_t),a_t,b_t,c_t,d_t\in \mathbb{R}, \forall t\in \mathcal{T}^h \}, \notag \\
	W_h &= \{ w_h \in W: w_h \ \text{is a piecewise constant function with respect to } \mathcal{T}^h \}. \notag
\end{align}
Then the discrete weak formulation is written as: Find $(\mathbf{u}_h,p_h)\in (V_h^0,W_h)$ such that
\begin{align}
	\int_\Omega \kappa^{-1}\mathbf{u}_h\cdot \mathbf{v_h} - \int_\Omega \text{div}(\mathbf{v}_h)p_h &= 0 ~~~~~~\forall\mathbf{v}_h\in V_h^0, \notag \\
	-\int_\Omega\text{div}(\mathbf{u}_h)w_h &= -\int_\Omega fw_h ~~~~~~\forall w_h \in W_h,
\end{align}
where 
$V_h^0=\{\mathbf{v}_h \in V_h: \mathbf{v}_h\cdot\mathbf{n}=0~\text{on}~\partial\Omega\}$. 

Let $\phi_1,\phi_2,\cdots,\phi_n$ and $w_1,w_2,\cdots,w_m$ be the basis of $V_h$ and $W_h$, respectively, where $m$ is the number of elements of $\mathcal{T}^h$ and $n$ is the number of edges of $\mathcal{T}^h$. Assume that
\begin{equation}
	\mathbf{u}_h=\sum_{i=1}^n u_i\phi_i,~~p_h=\sum_{j=1}^mp_jw_j, \notag
\end{equation}
then the matrix representation can be written as:
\begin{equation}
	\begin{bmatrix}
		M & B^T \\
		B & O
	\end{bmatrix}
	\begin{bmatrix}
		\mathbf{u}_h \\ p_h
	\end{bmatrix}
	=
	\begin{bmatrix}
		O \\ F
	\end{bmatrix}. 
	\label{saddle_system}
\end{equation}
where $M$ is a symmetric positive definite matrix, whose entries $M_{ij}:=\int_{\Omega}\kappa^{-1}\phi_i\phi_j$, $B$ is an approximation to the divergence operator with $B_{ij}:=-\int_{\Omega}w_i\nabla\cdot \phi_j$, and $F$ is a vector with $F_i := -\int_{\Omega}fw_i$.

By using the trapezoidal quadrature rule for the computation of $\int_{\Omega}\kappa^{-1}\phi_i\phi_j$, $M$ turns into a diagonal matrix and is easy to invert, so that we can solve \eqref{saddle_system} in the following way:
\begin{equation}
	(BM^{-1}B^T)p_h = F,~~\mathbf{u}_h = -M^{-1}(B^Tp_h). \notag
\end{equation}

Thanks to the above velocity elimination, we only need to establish the multiscale sapce for pressure. Suppose that $\{\psi^i_k\}$ is the set of local multiscale basis functions for the coarse element $T_i\in\mathcal{T}^H$, then the multiscale space for pressure is defined as: 
\begin{equation}
	W_H=\bigoplus_{T_i\in\mathcal{T}^H} \text{span}\{\psi^i_k\}. \notag
\end{equation}
Accordingly, the mixed GMsFEM is to find $(\mathbf{u}_H,p_H)\in (V_h,W_H)$ such that
\begin{align}
	\int \kappa^{-1}\mathbf{u}_H\cdot \mathbf{v}_H - \int \text{div}(\mathbf{v}_H)p_H &= 0 ~~~~~~\forall \mathbf{v}_H \in V_h^0, \notag \\
	-\int \text{div}(\mathbf{u}_H)w_H &= -\int fw_H ~~~~~~\forall w_H \in W_H.
\end{align} 

\subsection{Multiscale Basis Functions}
We first construct the local snapshot space $W_{\text{snap}}^{(i)}$ for each coarse element $T_i\in\mathcal{T}^H$. Let $J_i$ be the number of boundary edges of the fine-grid partition of $T_i$. There are three different ways: 
\begin{enumerate}
	\item The local snapshot space is taken as the fine-grid space restricted to $T_i$: 
	\begin{equation}
		W_{\text{snap}}^{(i)}=W_h(T_i). \notag
	\end{equation}
	\vspace{-5pt}
	
	\item Solve the local problem with Dirichlet boundary conditions:
	\begin{equation}
		\begin{cases}
			\kappa^{-1}\phi_j^{(i)}+\nabla \psi_j^{(i)}=0~~~\text{in}~T_i, \\ 
			\text{div}(\phi_j^{(i)})=0~~~\text{in}~T_i, \\ 
			\psi_j^{(i)}=\delta_j^{(i)}=
			\begin{cases}
				1~~\text{on}~e_j\in\partial T_i, \\
				0~~\text{on other fine-grid edges of}~\partial T_i,
			\end{cases}~j=1\cdots J_i.
		\end{cases}
		\label{way_2}
	\end{equation}
	Then the local snapshot space can be constructed as: 
	\begin{equation}
		W_{\text{snap}}^{(i)}=\text{span}\{\psi^{(i)}_j:j=1,2,\cdots,J_i\}. \notag
	\end{equation}
	\item Solve the local problem with Neumann boundary conditions:
	\begin{equation}
		\begin{cases}
			\kappa^{-1}\phi_j^{(i)}+\nabla \psi_j^{(i)}=0~~~\text{in}~T_i, \\
			\text{div}(\phi_j^{(i)})= \alpha^{(i)}_j~~~\text{in}~T_i, \\
			\frac{\partial \psi_j^{(i)}}{\partial \mathbf{n}_i} = \delta_j^{(i)}~~~\text{on}~\partial T_i,~~~j=1\cdots J_i,
		\end{cases}
		\label{way_3}
	\end{equation}
	where $\mathbf{n}_i$ is the outward unit normal vector on $\partial T_i$, $\alpha^{(i)}_j$ is chosen such that the compatibility condition $\int_{T_i}\alpha^{(i)}_j = \int_{\partial T_i}\delta^{(i)}_j$ is satisfied. In the same manner, 
	\begin{equation}
		W_{\text{snap}}^{(i)}=\text{span}\{\psi^{(i)}_j:j=1,2,\cdots,J_i\}. \notag
	\end{equation}
\end{enumerate}

The above problems (\ref{way_2}) and (\ref{way_3}) are solved numerically on the fine-grid partition of $T_i$ and using the $\text{RT}_0$ mixed finite element method such that $\psi_j^{(i)}\in W_h$.

Next, we construct the multiscale basis functions. For each local snapshot space $W_{\text{snap}}^{(i)}$, we reduce the spatial dimension by a local spectral problem \cite{chen2020generalized}:

\begin{equation}
	\begin{cases}
		a_{i}(\psi,w)=\lambda s_i(\psi,w) ~~~~\forall w\in W^{(i)}_{\text{snap}}, \\
		a_i(\psi,w)=\sum_{e\in \mathcal{E}_h^0}\kappa_e\left[\psi\right]_e\left[w\right]_e, \quad s_i(\psi,w)=\int_{T_i}\kappa \psi w, 
	\end{cases}
	\label{lsp}
\end{equation}
where $\mathcal{E}_h^0$ refers to the set of all interior edges of the fine-grid partition of $T_i$, $\kappa_e$ denotes the harmonic average of the permeability $\kappa$ on the two fine elements that have the common edge $e$, $[\psi]_e$ and $[w]_e$ are the jump of $\psi$ and $w$ across the edge $e$, respectively. The matrix form can be written as:
\begin{equation}
	A_iZ_k=\lambda_kS_iZ_k,
\end{equation}
where $A_i$ and $S_i$ are the stiffness matrix and mass matrix, respectively. $\lambda_k$ refers to the eigenvalue and $Z_k$ is the corresponding eigenvector. The eigenvalues are arranged in ascending order, and the first $l_i$ eigenvectors corresponding to the smallest eigenvalues are selected to generate the multiscale basis functions (offline basis functions):
\begin{equation}
	\psi^{i,\text{off}}_k=\sum_{j=1}^{J_i}Z_{k,j}\psi_j^{(i)},\quad k=1,2,\cdots,l_i,
\end{equation}
where $Z_{k,j}$ represents the $j$th component of $Z_k$. Then the offline basis functions of all coarse elements are combined to generate the multiscale space:
\begin{equation}
	W_H=\bigoplus_{T_i\in\mathcal{T}^H} \text{span}\{\psi^{i,\text{off}}_k\}.
\end{equation}

The mixed GMsFEM maps features of the fine-grid scale onto the coarse-grid scale, enabling to accurately capture the heterogeneity of porous media while ensuring the computational efficiency and precision when modeling large-scale problems.

\section{Methodology}
\label{method}

\subsection{Data Acquisition}
In this section, we will introduce how we generalize random high-contrast permeability fields and describe the dataset used in our experiment.
\subsubsection{Karhunen-Lo\`{e}ve Expansion}
Karhunen-Lo\`{e}ve Expansion (KLE) is a spectral decomposition method widely used to generate stochastic fields with spatially correlated heterogeneity \cite{fukunaga1970application}. Unlike traditional random sampling, KLE provides a low-dimensional representation of the field by truncating high-order modes while preserving dominant statistical features.

For a log-normal permeability field $\kappa(x;w)$, we define its logarithmic transform $\mathbb{Z}(x;w)=\log \kappa(x;w)$, modeled as a Gaussian random field with mean $\mathbb{E}(x)$ and covariance kernel $\mathbb{C}_{\mathbb{Z}}(x,y)$. The KLE decomposes $\mathbb{Z}(x;w)$ as:
\begin{equation}
	\mathbb{Z}(x;w)=\mathbb{E}(x)+\sum_{i=1}^\infty \sqrt{\lambda_i} \phi_i(x)\xi_i(w),
\end{equation}
where $\lambda_i$ and $\phi_i$ are eigenvalues and eigenfunctions of the integral equation
\begin{equation}
	\int_{\Omega} \mathbb{C}_{\mathbb{Z}}\phi_i(y)dy=\lambda_i\phi_i(x),
\end{equation}
$\xi_i(w)$ are uncorrelated standard Gaussian random variables.

The generation of random fields follows the following steps:
\begin{enumerate}
	\item Define $\mathbb{C}_{\mathbb{Z}}$ to encode spatial correlation length $l$ and variance $\sigma^2$.
	\item Solve the discretized eigenproblem numerically.
	\item Retain the first $N_{\text{KLE}}$ modes, ensuring $\sum_{i=1}^N \lambda_i / \sum_{i=1}^\infty \lambda_i \geq 95\%$.
	\item Generate realizations by sampling $\xi_i \sim N(0,1)$ and computing $\mathbb{Z}(x;w)$, then exponentiating to obtain $\kappa(x;w)$.
\end{enumerate}

\subsubsection{Dataset}
We use KLE to generate the high-contrast permeability fields, and compute the multiscale basis functions using mixed GMsFEM based on the permeability fields. Set the computational domain $\Omega=\left[ 0,1 \right]^2$, and the value of the basis function between $-1$ and $1$. The fine grid is configured as a $30\times 30$ uniform mesh, while the coarse grid is set as a $10 \times 10$ uniform mesh, indicating that each coarse element contains $3 \times 3$ fine elements. Within the two-scale grid, there is a probability that a coarse element may contain fractures either completely or partially, and these fractures can intersect with one another.

\begin{figure}[h]
	\centering
	\includegraphics[width=0.6\linewidth]{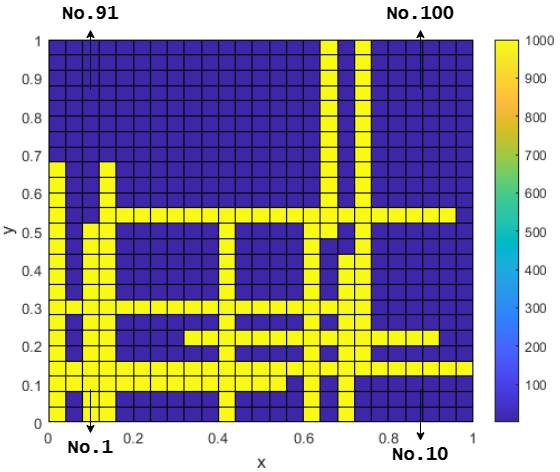}
	\caption{A permeability field of fractured porous media with $\kappa_m$ of 1 millidarcy and $\kappa_f$ of 1000 millidarcy. The number of fractures is 15. The domian in blue refers to the matrix, and those in yellow refer to the fractures. We assign coordinates to the coarse grid system from bottom to top and from left to right. The coarse element in the lower left corner is No.1, and so on.}
	\label{fig_perm example}
\end{figure}

For parameter setting, the matrix permeability $\kappa_m$ has 5 different values: {1, 2, 3, 4, 5} millidarcy, while the fracture permeability $\kappa_f$ has 7 different values: {500, 750, 1000, 1250, 1500, 1750, 2000} millidarcy. The number of fractures in the porous media ranges from 1 to 25, resulting in a total of 25 different configurations. This parameter setup yields 875 unique combinations ($5 \times 7 \times 25$). Using MATLAB code, we repeat generated random samples, producing a total of 177800 samples. Considering the randomness in the generation process, which could lead to duplicate samples, we performed a duplication-removing process and removed 6537 duplicates. As a result, we obtained a dataset suitable for neural network training, consisting of 102757 training samples, 34252 validation samples, and 34254 testing samples, maintaining a 6: 2: 2 ratio. These data will be used for feature learning and evaluation during the model training process, as well as to assess model performance after training completion.

Figure.\ref{fig_perm example} illustrates a sample of the permeability field, where the matrix permeability $\kappa_m$ is 1 millidarcy, and the fracture permeability $\kappa_f$ is 1000 millidarcy, with a total of 15 fractures. We sort the coarse element in the bottom left corner of the figure as No.1 in coordinates that increase from left to right and from bottom to top. In this multiscale system, the basis functions are defined on the coarse element, but elements with different coordinates may have distinct basis functions even under the same permeability conditions (for instance, coarse element No.10 and No.91 in Figure.\ref{fig_perm example}).

The initial permeability field was represented as a $900 \times 1$ one-dimensional tensor. We reconstructed it into a $100 \times 9$ two-dimensional tensor, where 100 and 9 correspond to the number of coarse elements and the number of fine elements within each coarse element, respectively, while the structure of the basis function remained unchanged. This transformation enables the use of two-dimensional multiscale convolutional filters for feature learning during the model construction. Consider that batch normalization was frequently applied within the model, data normalization was not performed during the preconditioning stage.

\subsection{Fourier transform-based Preconditioner}

\begin{figure}[t]
	\centering
	\includegraphics[width=1.0\linewidth]{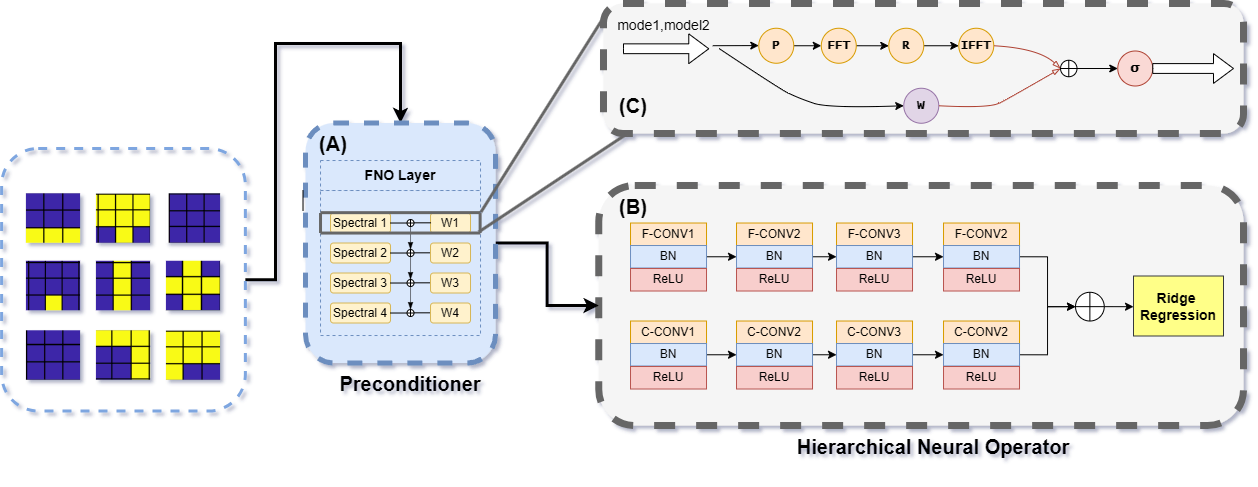}
	\caption{Architecture of our proposed Preconditioner-Learner. (A) Fourier transform-based preconditioner. 'Spectral' denotes spectral convolution. (B) Hierarchical neural operator with two-grid filters. (C) Structure of a single spectral convolution layer in (A).}
	\label{fig_network structure}
\end{figure}

FNO is the first step of our model, acting as a preconditioner, transforming the input $x$ into a richer representation in the frequency domain. The preconditioner applies the Fourier integral operator $\mathcal{K}$ to convert $x$ into a spectral representation, which enables the model to capture local and global dependencies, giving a full understanding of the spatial relationships inherent in the basis functions. The Fourier integral operator is defined as:
\begin{equation}
	(\mathcal{K}(\phi)v_t)=\mathcal{F}^{-1}(R_{\phi}\cdot (\mathcal{F}v_t)(x))\quad \forall x\in D,
	\label{eq_Fourier integral operator}
\end{equation}
where $R_{\phi}$ is the Fourier transform of a periodic function $k:\bar{D}\rightarrow \mathbb{R}^{d_v \times d_v}$, parameterized by $\phi \in \Theta_{\mathcal{K}}$.

The operator $\mathcal{K}$ plays a central role in converting the permeability field to the frequency domain, thus achieving efficient spectral convolution. By learning $R_{\phi}$ the model adapts to different frequency component of the input, effectively capturing the global interaction, mapping it to the frequency domain:
\begin{equation}
	(\mathcal{F}v)(\xi) = \langle v,\psi(\xi) \rangle_{L(D)} = \int_x v(x)\psi(x,\xi)\mu(x) \approx \sum_{x\in \mathcal{T}}v(x)\psi(x,\xi),
\end{equation}
where $\psi(x,\xi)=\exp(2\pi i \langle x,\xi \rangle ) \in L(D)$ is the Fourier basis function and $\mathcal{T}$ is the mesh sampled from distribution $\mu$. $\xi \in D$ is the frequency mode. Assume that $D$ is a periodic, square torus and the mesh $\mathcal{T}$ is uniform, then the Fourier transform $\mathcal{F}$ can be implemented by the 2D fast Fourier transform (2D-FFT):
\begin{equation}
	\hat{x}(\xi)=\int_{\Omega} xe^{-2\pi i \langle x,\xi \rangle}dx.
\end{equation}

In the frequency domain, the convolution operation is transformed into element-wise multiplication, which is efficient in computation:
\begin{equation}
	\hat{y}=\hat{\mathcal{K}}(\xi)\cdot \hat{x}(\xi),
\end{equation}
where $\hat{\mathcal{K}}(\xi_1,\xi_2)$ represents the learnable kernel in the frequency domain, which allows the necessary operations in the domain. Then the modified data is mapped back to the spatial domain using the inverse Fourier transform (IFFT):
\begin{equation}
	y(x)=\int_{\xi} \hat{y}(\xi)e^{2\pi i \langle x,\xi \rangle}d\xi,
\end{equation}
this step enables the model to recover spatial patterns while taking advantages of spatial learning.

The process of FNO layers reconstruct the input in the original domain and enrich the learning spectral information to capture global and local dependencies.

Our FNO-based preconditioner consists of multiple spectral convolution layers (here we set 4 layers). Figure.\ref{fig_network structure}-(A)(B) demonstrate the whole structure combining all the theories mentioned above, and convert the permeability field from spatial domain to frequency domain, then back to the spatial domain after pattern trimming. Each layer can capture different levels of frequency information. For each level, the process the input goes through can be written as Algorithm.\ref{alg_fno_preconditioner}:

\begin{algorithm}[h]
	\caption{Pseudocode of FNO-based Preconditioner}
	\label{alg_fno_preconditioner}
	\begin{algorithmic} 
		\REQUIRE High-contrast Permeability Field $\kappa(x)$, Parameters $\phi, R, W$
		\ENSURE Preconditioned field $\kappa_{\text{output}}(x)$
		
		\STATE \textbf{Lifting and Fourier Transform:}
		\STATE \quad Map $\kappa(x)$ to a higher dimension
		\STATE \quad $\hat{\kappa}(\xi) \gets \mathcal{F}(\kappa(x))$ \COMMENT{$\mathcal{F}$ is Fourier transform}
		
		\STATE \textbf{Linear Transformation on Frequency Domain:}
		\STATE \quad Filter higher frequency modes
		\STATE \quad $\hat{\kappa}_{\text{new}}(\xi) \gets R(\xi) \cdot \hat{\kappa}(\xi)$
		
		\STATE \textbf{Inverse Fourier Transform:}
		\STATE \quad $\kappa_{\text{new}}(x) \gets \mathcal{F}^{-1}(\hat{\kappa}_{\text{new}}(\xi))$
		
		\STATE \textbf{Local Linear Projection:}
		\STATE \quad $\kappa_{\text{local}}(x) \gets W(\kappa(x))$ \COMMENT{Point-wise linear transform}
		
		\STATE \textbf{Feature Fusion:}
		\STATE \quad $\kappa_{\text{output}}(x) \gets \sigma\big( \kappa_{\text{new}}(x) + \kappa_{\text{local}}(x) \big)$ \COMMENT{$\sigma$ is GELU activation function}
	\end{algorithmic}
\end{algorithm}

\subsection{Miltiscale Feature Extraction and Fusion}

Deep learning techniques have superior performance in constructing PDEs and basis functions solvers. A conventional CNN consists of input, convolution, pooling, fully connected (FC), and output layers. The more complex the model has a larger order of parameters, which will put a certain demand on the computational resource and hardware. Compared to traditional networks used in image recognition, like AlexNet \cite{krizhevsky2012imagenet} and VGGNet \cite{simonyan2014very}, our multiscale neural operator constructed in this study uses fewer layers, has less complexity, and can obtain ideal results with lower time costs.

Although the basis functions we compute are defined on the coarse-grid system, the fine-grid system can still contain certain information. For this purpose, we employ convolution filters at two different scales for information learning. Considering that a single-path neural network structure may alter feature scales, using convolution kernels of two different scales within a single path could lead to the loss of some information. Hence, we have constructed a dual-path multiscale CNN architecture.

In this design, the output of the preconditioner is respectively passed through a large-scale net ($3 \times 3$) and a small-scale net ($1 \times 1)$, followed by feature fusion of the outputs from both paths, enabling multiscale learning and hierarchical pattern learning. Since the size of filter used by small-scale CNN is $1 \times 1$, this path does not change the specification and numerical results of data, then it can be regarded as a channel expansion operation of the initial information. This approach is pretty similar to the feature combination in FNO above, which enables maximum learning of the feature information constrained in the multiscale system.

In both the coarse and fine grid paths, the data output from the preconditioner is processed through a series of convolution layers. A nonlinear mapping is performed in each convolution layer:
\begin{equation}
	x_{\text{coarse or fine}}^{(i+1)} = \sigma(\text{BN}(W^{(i)}\cdot x_{\text{coarse or fine}}^{(i)})),
\end{equation}
where $W$ denotes the convolution filters with kernel size $3 \times 3$ or $1 \times 1$. $\text{BN}$ represents batch normalization, and $\sigma$ is the ReLU activation function. These layers capture spatial dependencies at both scales, effectively learning the multiscale features of the input. Then this process will be flattened and passed to a FC layer. This layer implements ridge regression using $L2$ regularization:
\begin{equation}
	y=W_{\text{ridge}}\cdot x_{\text{flattened}}+b,
\end{equation}
where $W_{\text{ridge}}$ is the learnable weights, and $x_{\text{flattened}}$ is the flattened feature map. The ridge regression loss function is given by:
\begin{equation}
	\mathcal{L}=\frac{1}{N}\sum_{i=1}^N (y_i-\hat{y}_i)^2+\lambda\|W_{\text{ridge}}\|_2^2,
\end{equation}
where $\lambda\|W_{\text{ridge}}\|_2^2$ is the $L2$ regularization term, encouraging the model to maintain smaller weights, thereby improving generalization and preventing overfitting.

The multiscale operator captures the hierarchical spatial structures in the data by learning both coarse and fine grid features through separate pathways, followed by channel fusion. \hyperref[fig_network structure]{Figure.\ref{fig_network structure}}-(B) illustrates the multiscale structure for learning multiscale basis functions. This approach allows the model to adapt to details at different scales, thereby enhancing its ability to generalize to solutions of different types of basis functions.

The complete model integrates the spectral representation from the FNO-based preconditioner with the multiscale convolution network. This integration enables the model to capture both global (frequency-based) and local (spatial-based) dependencies, which is crucial for effectively modeling and solving basis function problems.

\subsection{Evaluation Metrics}
The evaluation of this research is divided into several stages to fully assess its performance and stability.

(\textbf{Model Performance}) Three standard metrics are selected to evaluate the accuracy of the predicted basis functions: mean squared error (MSE), mean absolute error (MAE), and coefficient of determination ($\text{R}^2$).

MSE quantifies the average squared difference between predicted and actual values, offering insight into overall prediction accuracy and penalizing larger errors more severely.
\begin{equation}
	\text{MSE}=\frac{1}{N}\sum_{i=1}^N(y_i-\hat{y}_i)^2
\end{equation}

MAE measures the average absolute different between predictions and true values, providing a clearer indication of the model's precision without over-penalizing outliers.
\begin{equation}
	\text{MAE}=\frac{1}{N}\sum_{i=1}^N|y_i-\hat{y}_i|
\end{equation}

$\text{R}^2$ represents the proportion of variance in the dependent variable that is explained by the independent variables, serving as an indicator of the quality of the model fit.
\begin{equation}
	\text{R}^2=1-\frac{\sum_{i=1}^N (y_i-\hat{y}_i)^2}{\sum_{i=1}^N (y_i -\bar{y})^2}
\end{equation}
where $y_i$, $\hat{y}_i$ and $\bar{y}$ denote the true value, predicted value and average of true values.

(\textbf{Stability Assessment}) To evaluate the model's stability and stability, perturbations are induced into the input data and subsequently re-evaluate the model using the same metrics. This allows us to observe how the model's performance changes when exposed to noise, thereby assessing its resilience to data variability. A stable model is expected to demonstrate minimal variation in these metrics when subjected to such perturbations, indicating its stability under non-ideal conditions. In addition, we also hired the learning curve as one of the metrics, which shows whether the training loss is converged.

\section{Analysis}
\label{theorem}

\subsection{Complexity Analysis}
In this section we will compute and compare the time and space complexity of our proposed method, machine learning-based method (CNN for instance), and mixed GMsFEM.

Our proposed model combines FNO with a multiscale neural network. With input size $N\times N$, the main steps of the model are:
\begin{enumerate}
	\item \textbf{FFT}: Applying the Fourier transform to the input data in the frequency domain has a time complexity of $\mathcal{O}(N^2 \log N)$, where $N\times N$ represents the size of the input grid.
	\item \textbf{Spectral Convolution}: The operation in the frequency domain is element-wise multiplication, which is $\mathcal{O}(N^2)$.
	\item \textbf{IFFT}: The inverse operation has the same complexity as FFT, i.e. $\mathcal{O}(N^2 \log N)$.
	\item The multiscale neural network applies convolutions using multiscale filters, which have a complexity of $\mathcal{O}(N^2)$ for each convolution layer. If there are $L$ layers, the overall complexity of this part is $\mathcal{O}(L\times N^2)$.
\end{enumerate}

Therefore, the overall time complexity of our model is:
\begin{equation}
	\mathcal{O}(N^2\log N)+\mathcal{O}(L\times N^2) \notag
\end{equation}

Since FFT and IFFT operations require storing the frequency-domain representations, the space complexity of these is $\mathcal{O}(N^2)$. The space complexity for storing the weights and activations of each convolution layer is $\mathcal{O}(L\times N^2$. Thus the overall space complexity of our proposed model is
\begin{equation}
	\mathcal{O}(N^2)+\mathcal{O}(L\times N^2)=\mathcal{O}(L\times N^2) \notag
\end{equation}
where $L$ is the number of layers.

For the traditional machine learning-based method, CNN, the time complexity for a single convolution operation is $\mathcal{O}(K^2 \times N^2)$, where $K$ is the filter size and $N$ is the input grid size. If there are $L$ convolution layers, the total time complexity is $\mathcal{O}(L \times K^2 \times N^2)$. After the convolution layers, the data is flattened and passed through a fully connected layer (FC layer). The time complexity for this is $\mathcal{O}(N^2 \times F)$, where $F$ is the numbers of neurons in the FC layer.

Thus, the overall time complexity of the 2D CNN is
\vspace{-5pt}
\begin{equation}
	\mathcal{O}(L \times K^2 \times N^2)+\mathcal{O}(N^2 + F) \notag
\end{equation}

The space complexity of 2D-CNN has been mentioned above. The space complexity for storing the weights in the FC layer is $\mathcal{O}(N^2 \times F)$. So the total space complexity of ML-based method is
\begin{equation}
	\mathcal{O}(L \times K^2 \times N^2)+\mathcal{O}(N^2 \times F) \notag
\end{equation}

For mixed GMsFEM, the time complexity consists of below components:
\begin{enumerate}
	\item \textbf{Solving PDEs}: Solving a local cell problem for each coarse grid requires solving a set of PDEs. If there are $N_c$ coarse grid elements, the time complexity for these is $\mathcal{O}(N_c \times N_f)$, where $N_f$ is the number of find-scale points per coarse grid.
	\item \textbf{Eigenvalue Problems}: For each coarse grid, eigenvalue problem need to be solved to extract the dominant modes. The time complexity for each eigenvalue problem is $N_f^3$, where $N_f$ is the fine grid edges within a coarse grid. Given $N_c$ coarse elements, the total time complexity is $\mathcal{O}(N_c \times N_f^3)$.
	\item \textbf{Matrix Assembly and Solution}: After the basis functions are computed, a system of equations must be solved. The time complexity for assembling and solving this system is $\mathcal{O}(N_c \times N_f^2)$.
\end{enumerate}

Therefore, the total time complexity of mixed GMsFEM is
\begin{equation}
	\mathcal{O}(N_c \times N_f^3) \notag
\end{equation}

The space complexity for storing the local problems is $\mathcal{O}(N_c \times N_f)$, and storing the eigenvectors from each eigenvalue problem is $\mathcal{O}(N_f^2)$, for storing the matrix for the system equations is $\mathcal{O}(N_c \times N_f^2)$.

So the overall space complexity of mixed GMsFEM is
\begin{equation}
	\mathcal{O}(N_c \times N_f^2) \notag
\end{equation}

\begin{obse}[Comparison of Different Methods]
    All time and space complexity of the mentioned three methods can be seen in the Table.\ref{tab_complexity_comparison}:
    \renewcommand{\arraystretch}{1.3}
	\begin{table}[h]
		\footnotesize
		\centering
		\caption{Comparison of Computational Complexity.}
		\setlength{\tabcolsep}{6pt} 
		\begin{tabular}{l p{4.5cm} p{4.5cm}}
			\toprule
			\textbf{Method} & \textbf{Time Complexity} & \textbf{Space Complexity} \\
			\midrule
			FP-HMsNet & \( \mathcal{O}(N^2 \log N) + \mathcal{O}(L \times N^2) \) & \( \mathcal{O}(L \times N^2) \) \\
			2D CNN \cite{choubineh2022innovative} & \( \mathcal{O}(L \times K^2 \times N^2) + \mathcal{O}(N^2 \times F) \) & \( \mathcal{O}(L \times K^2 \times N^2) + \mathcal{O}(N^2 \times F) \) \\
			Mixed GMsFEM \cite{chen2020generalized} & \( \mathcal{O}(N_c \times N_f^3) \) & \( \mathcal{O}(N_c \times N_f^2) \) \\
			\bottomrule
		\end{tabular}
		\label{tab_complexity_comparison}
	\end{table}

\begin{enumerate}
	\item The model we proposed relies on frequency domain transformation and low-rank linear mapping, achieving the best computational efficiency and lightweight parameters among the three methods we mentioned. It uses frequency domain efficiency to significantly outperform spatial methods in high-resolution scenarios. When the resolution doubles, its computational growth is better than that of linear models, making it suitable for high-frequency or high-resolution signal processing (such as fluid simulation) and some memory-sensitive scenarios.
	\item For ML-based methods, affected by the nature of the convolution operation, it consumes a lot of memory to store a large number of redundant parameters. This method is suitable for capturing local patterns, but the complexity grows with the square of the resolution, so it is not sustainable in high-resolution scenarios and is only suitable for low-resolution image/speech processing and end-to-end training that does not require prior knowledge in the frequency domain.
	\item The core advantage of mixed GMsFEM lies in its ability to preprocess basis functions offline and solve quickly online in multiscale modeling. This method is very suitable for ultra-large-scale physical simulation. In addition, this numerical method is not similar to a black box and has an advantage in interpretability. This method is acceptable when offline preprocessing is acceptable.
\end{enumerate}
\end{obse}

By replacing spatial convolutions with FFT-based spectral operations, we achieve time complexity of $\mathcal{O}(N^2\log N)$, enabling real-time simulations on $10^6$-scale grids. The $\mathcal{O}(LN^2)$ memory footprint avoids the parameter explosion of CNN, making our method deployable on memory-constrained devices.

\subsection{Error, Stability and Convergence}
In this section we will analysis the error, stability, and convergence of our proposed model, providing corresponding theorems to demonstrate the properties.

\begin{lem}[Lipschitz Continuity of Neural Operator \cite{kovachki2023neural,shang2022lipschitz}]
	Let 
    \begin{equation}\mathcal{N}: L^2(\Omega) \mapsto H^1(\Omega)  \notag \end{equation} 
    denote the multiscale neural operator. Assume the network weights are bounded, then there exists a constant $L_{\mathcal{N}}>0$ such that:
	\begin{equation}
		\| \mathcal{N}(z_1)-\mathcal{N}(z_2)\|_{H^1} \leq L_{\mathcal{N}}\|z_1 - z_2\|_{L^2},\ \forall \ z_1, z_2 \in L^2(\Omega)
	\end{equation}
	\label{lem_continuity_neural_operator}
\end{lem}

\begin{theo}[Error Bound of FP-HMsNet]
	Given the permeability $\kappa$ and corresponding multiscale basis function $\psi$, the output of FP-HMsNet, denoted as $\hat{\psi}$. $\mathcal{F}$ and $\mathcal{N}$ are the true operators of preconditioner and multiscale neural operator, respectively. And $\hat{\mathcal{F}}$ and $\hat{\mathcal{N}}$ are the learned operators of $\mathcal{F}$ and $\mathcal{N}$. Assume that $\mathcal{N}$ is $L_{\mathcal{N}}$-Lipschitz continuous, $\forall~\varepsilon_{\mathcal{N}},\varepsilon_{\mathcal{F}}>0$, the error of FP-HMsNet satisfies:
	\begin{equation}
		\| \psi-\hat{\psi}\|_{L^2} \leq C_1\varepsilon_{\mathcal{F}}+C_2\varepsilon_{\mathcal{N}}
	\end{equation}
	where $C_1$ is a constant dependent on $\mathcal{N}$, $C_2$ is a constant independent of $\mathcal{F}$ and $\mathcal{N}$.
	\label{th_error_bound}
\end{theo}

\textbf{Proof}:FP-HMsNet consosts of two components:
	\begin{enumerate}
		\item FNO-based preconditioner: $\kappa\mapsto\mathcal{F}(\kappa)$
		\item Multiscale neural operator: $\mathcal{F}(\kappa)\mapsto\psi$
	\end{enumerate}
	
	The error of FP-HMsNet can be written as:
	\begin{equation}
		\psi-\hat{\psi}=\mathcal{N}(\mathcal{F}(\kappa))-\hat{\mathcal{N}}(\hat{\mathcal{F}}(\kappa)) \notag
	\end{equation}

Take $H^1$-norm, by triangle inequality we have:
	\begin{align}
		\| \psi - \hat{\psi} \|_{H^1} &= 
		\| \mathcal{N}(\mathcal{F(\kappa}))-\mathcal{N}(\hat{\mathcal{F}}(\kappa))+\mathcal{N}(\hat{\mathcal{F}}(\kappa)) - \hat{\mathcal{N}}(\hat{\mathcal{F}}(\kappa))\|_{H^1} \notag \\ &\leq 
		\underbrace{\| \mathcal{N}(\mathcal{F}(\kappa))-\mathcal{N}(\hat{\mathcal{F}}(\kappa))\|_{H^1}}_{\text{Term 1}} + \underbrace{\| \mathcal{N}(\hat{\mathcal{F}}(\kappa)) - \hat{\mathcal{N}}(\hat{\mathcal{F}}(\kappa)) \|_{H^1}}_{\text{Term 2}}
	\end{align}
	
	(\textbf{Term 1}) By Lemma.\ref{lem_continuity_neural_operator}, there exists a Lipschitz constant $L_{\mathcal{N}}$ such that:
	\begin{equation}
		\| \mathcal{N}(\mathcal{F}(\kappa))-\mathcal{\mathcal{N}}(\hat{\mathcal{F}}(\kappa))\|_{H^1} \leq L_{\mathcal{N}}\|\mathcal{F}(\kappa)-\hat{\mathcal{F}}(\kappa)\|_{L^2}
	\end{equation}
	
	By Sobolev embedding theorem, there is a embed constant $C_{\text{embed}}$ such that:
	\begin{equation}
		\| \cdot \|_{L^2} \leq C_{\text{embed}}\| \cdot \|_{H^1}
		\label{eq_sobolev_embed}
	\end{equation}
	
	By the universal approximation theorem of neural operator \cite{cybenko1989approximations,hornik1991approximation}, $\forall~\varepsilon_{\mathcal{F}>0}$, we can derive from \eqref{eq_sobolev_embed} that
	\begin{equation}
		\| \mathcal{N}(\mathcal{F}(\kappa))-\mathcal{N}(\hat{\mathcal{F}}(\kappa))\|_{H^1} \leq C_{\text{embed}}L_{\mathcal{N}}\|\mathcal{F}(\kappa)-\hat{\mathcal{F}}(\kappa)\|_{H^1} \leq C_{\text{embed}}L_{\mathcal{N}}\varepsilon_{\mathcal{F}}
		\label{eq_error_term1}
	\end{equation}

    (\textbf{Term 2}) By the universal approximation theorem of neural operator, $\forall\ \varepsilon_{\mathcal{N}}>0$, the norm satisfies the following inequality:
	\begin{equation}
		\| \mathcal{N}(\hat{\mathcal{F}}(\kappa))-\hat{\mathcal{N}}(\hat{\mathcal{F}}(\kappa))\|_{H^1} \leq \varepsilon_{\mathcal{N}}
		\label{eq_error_term2}
	\end{equation}
	
	Combining Equation.\eqref{eq_error_term1} and Equation.\eqref{eq_error_term2} and take $L^2$-norm of error:
	\begin{align}
		\|\psi - \hat{\psi}\|_{L^2} &\leq 
		C_{\text{embed}}\|\psi-\hat{\psi}\|_{H^1} \notag \\ &\leq
		C_{\text{embed}}(C_{\text{embed}}L_{\mathcal{N}}\|\mathcal{F}(\kappa)-\hat{\mathcal{F}}(\kappa)\|_{H^1}+\varepsilon_{\mathcal{N}}) \notag \\ &=C_1\varepsilon_{\mathcal{F}}+C_2\varepsilon_{\mathcal{N}}
	\end{align}
	where $C_1=C_{\text{embed}}^2L_{\mathcal{N}}$, $C_2=C_{\text{embed}}$. \solidqed

\begin{theo}[Stability of FP-HMsNet]
	Let $\hat{\psi}$ denote the multiscale basis function derived from FP-HMsNet, respectively. Then $\forall~\varepsilon>0$ and there exists a constant $L_{\text{global}}$ such that the difference of predicted multiscale basis functions comes mostly from the difference of the input permeability fields, i.e. $\forall~\kappa_1,\kappa_2$ we have 
	\begin{equation}
		\| \hat{\psi}(\kappa_1)-\hat{\psi}(\kappa_2)\|_{H^1(\Omega)} \leq 2\varepsilon+L_{\text{global}}\sum_{T_i \in \mathcal{T}^H}\| \kappa_1 - \kappa_2\|_{L^{\infty}(T_i)}
	\end{equation}
	\label{th_stability} 
\end{theo}

\textbf{Proof}: 
	According to the local spectral problem (\ref{lsp}), when the permeability changes from $\kappa_1$ to $\kappa_2$, the perturbation of $a_i(\cdot,\cdot)$ is
	\begin{equation}
		\delta a_i(p,w)=\int_{T_i}(\kappa_1-\kappa_2)\nabla p \cdot \nabla w
	\end{equation}
	
	By the perturbation theory of linear operator \cite{kato2013perturbation}, the change of eigenfunctions satisfies:
	\begin{equation}
		\| \psi^{i,\text{off}}(\kappa_1)-\psi^{i,\text{off}}(\kappa_2)\|_{H^1(T_i)} \leq \frac{\| \kappa_1-\kappa_2\|_{L^{\infty}(T_i)}}{\kappa_{\text{min}}}\|\nabla \psi^{i,\text{off}}(\kappa_2)\|_{L^2(T_i)}
	\end{equation}
	where $\psi^{i,\text{off}}(\kappa)$ is the offline basis function of $\kappa$ on the coarse element $T_i$.
	
	Considering the energy estimation of spectral problems $\| \nabla \psi^{i,\text{off}}(\kappa)\|_{L^2} \\ \leq \sqrt{\lambda^{(i)}_{\text{max}}}\|\psi^{i,\text{off}}\|_{L^2}$\cite{gilbarg1977elliptic} and the normalization conditions $\| \psi^{i,\text{off}}(\kappa)\|_{L^2}=1$, we have
	\begin{equation}
		\| \psi^{i,\text{off}}(\kappa_1)-\psi^{i,\text{off}}(\kappa_2)\|_{H^1(T_i)} \leq L_{\text{local}}\|\kappa_1 - \kappa_2\|_{L^{\infty}(T_i)},~~ L_{\text{local}}=\frac{\sqrt{\lambda_{\text{max}}^{(i)}}}{\kappa_{\text{min}}}
	\end{equation}
	where $\kappa_\text{min}=\min \kappa,~~\kappa_{\text{max}}=\max \kappa~~, \kappa_{\text{min}}<\kappa < \kappa_{\text{max}}$.

    The difference of basis functions can be decomposed as:
	\begin{equation}
		\| \psi(\kappa_1)-\psi(\kappa_2)\|_{H^1(\Omega)}^2 = \sum_{T_i \in \mathcal{T}^H} \|\psi^{i,\text{off}}(\kappa_1)-\psi^{i,\text{off}}(\kappa_2)\|_{H^1(T_i)}^2
	\end{equation}
	
	For every single term in the right side of above inequality, we have 
	\begin{equation}
		\| \psi^{i,\text{off}}(\kappa_1)-\psi^{i,\text{off}}(\kappa_2)\|_{H^1(T_i)}^2 \leq L^2_{\text{local}}\|\kappa_1-\kappa_2\|^2_{L^{\infty}(T_i)}
	\end{equation}
	
	Sum up for all coarse elements and let $L_{\text{global}}=\sqrt{\sum_{i=1}^{N_e}(L^{(i)}_{\text{local}^2})}$, we can obtain
	
	\begin{align}
		\| \psi(\kappa_1)-\psi(\kappa_2)\|_{H^1(\Omega)}^2 &=
		\sum_{T_i \in \mathcal{T}^H}\| \psi^{i,\text{off}}(\kappa_1)-\psi^{i,\text{off}}(\kappa_2)\|_{H^1(T_i)}^2 \notag \\ &\leq
		\sum_{T_i \in \mathcal{T}^H}\left(L^{(i)}_{\text{local}}\right)^2 \| \kappa_1 - \kappa_2 \|^2_{L^{\infty}(T_i)} \\ & \leq
		\left(\sum_{T_i \in \mathcal{T}^H}L_{\text{global}}\| \kappa_1 - \kappa_2\|_{L^{\infty}(T_i)}\right)^2 \notag
	\end{align}
	
	Thus we have
	\begin{equation}
		\| \psi(\kappa_1)-\psi(\kappa_2)\|_{H^1(\Omega)} \leq
		L_{\text{global}}\sum_{T_i \in \mathcal{T}^H}\| \kappa_1 - \kappa_2\|_{L^{\infty}(T_i)}
	\end{equation}
	
	Take $H^1$-norm of the difference of $\hat{\psi}$ and by triangle inequality we have
	\begin{multline}
		\| \hat{\psi}(\kappa_1)-\hat{\psi}(\kappa_2)\|_{H^1(\Omega)} \\
		=\| (\hat{\psi}(\kappa_1)-\psi(\kappa_1)) - (\hat{\psi}(\kappa_2)-\psi(\kappa_2)) + (\psi(\kappa_1)-\psi(\kappa_2))\|_{H^1(\Omega)} \\
		\leq \| \underbrace{\hat{\psi}(\kappa_1)-\psi(\kappa_1)}_{\text{Term 1}}\|_{H^1(\Omega)} + \| \underbrace{\hat{\psi}(\kappa_2)-\psi(\kappa_2)}_{\text{Term 2}}\|_{H^1(\Omega)} \\
		+ \| \underbrace{\psi(\kappa_1)-\psi(\kappa_2)}_{\text{Term 3}}\|_{H^1(\Omega)}
		\label{eq_th_stability}
	\end{multline}
	
	For Term 1 and 2, from the proof of Theorem.\ref{th_error_bound}, $\forall~\varepsilon>0$ we have
	\begin{equation}
		\| \hat{\psi}(\kappa_i)-\psi(\kappa_i)\|_{H^1} \leq \varepsilon,~ i=1,2 \notag 
	\end{equation}

    Then we can rewrite \eqref{eq_th_stability} as
	\begin{equation}
		\| \hat{\psi}(\kappa_1)-\hat{\psi}(\kappa_2) \|_{H^1} \leq 2\varepsilon+L_{\text{global}}\sum_{T_i \in \mathcal{T}^H}\| \kappa_1 - \kappa_2\|_{L^{\infty}(T_i)}
	\end{equation}
    \solidqed
	
	This inequality implies that the difference of learned multiscale basis functions comes mostly from the difference of the input permeability fields, not from other instability.

Then we can derive the convergence of our model:
\begin{coro}[\textbf{Convergence}]
	Under the stability guarantee of FP-HMsNet and the universal approximation conditions, as the training set size $N\rightarrow \infty$ and the number of preserved Fourier modes $N_{\text{modes}}\rightarrow \infty$, the predicted multiscale basis function $\hat{\psi}$ converge to the true solution $\psi$ in the $L^2(\Omega)$-norm with probability 1. Specifically,
	
	\begin{equation}
		\lim_{N,N_\text{modes}\rightarrow \infty}\mathbb{E}_{\kappa \sim \mathcal{D}}\left[ \|\psi-\hat{\psi}\|_{L^2(\Omega)}\right]=0
	\end{equation}
	\label{coro_convergence}
\end{coro}

\textbf{Proof}: 
	Let the training set $\mathcal{D}$ be sampled from true distribution. By \hyperref[th_stability]{Theorem.\ref{th_stability}}, the model's output is Lipschitz continuous to input perturbations. Then the empirical error on $\mathcal{D}$ converges uniformly to the population error as $N \rightarrow \infty$:
	\begin{equation}
		\sup_{\kappa \in \mathcal{D}} \|\hat{\psi}-\psi\|_{L^2} \xrightarrow{N \to \infty} 0
	\end{equation}
	
	By the universal approximation theorem \cite{cybenko1989approximations}, for any $\varepsilon>0$, there exists a neural operator and corresponding parameterization $\theta^*$ such that:
	\begin{equation}
		\|\mathcal{N}(\mathcal{F}(\kappa))-\hat{\mathcal{N}}(\mathcal{F}(\kappa);\theta^*)\|_{H^1} \leq \varepsilon_{\mathcal{N}}
	\end{equation}
	
	As the training set size $N\rightarrow \infty$, the empirical risk minimization ensures $\varepsilon_{\mathcal{N}}\rightarrow 0$. The stability ensures the approximation is preserved under finite dataset size.
	
	The preconditioner $\mathcal{F}$ maps $\kappa$ to a low-dimensional spectral representation. From the spectral gap assumption, the preconditioner's truncation error \cite{adler2009random} decays exponentially:
	\begin{equation}
		\varepsilon_{\mathcal{F}}\leq C\cdot e^{-\gamma N_{\text{modes}}}
	\end{equation}
	where $N_{\text{modes}}$ is the number of preserved Fourier modes. As $N_{\text{modes}} \rightarrow \infty$, $\varepsilon_{\mathcal{F}}\rightarrow 0$.
	
	Substituting above results into the error bound (Theorem.\ref{th_error_bound}), we can obtain:
	\begin{equation}
		\| \psi - \hat{\psi} \|_{L^2} \leq C_1e^{-\gamma}N_{\text{modes}}+C_2\varepsilon_{\mathcal{N}}
	\end{equation}

As $N_{\text{modes}}\rightarrow \infty$ and $N \rightarrow \infty$, which means sufficient spectral resolution and data, both terms vanish, i.e.
	\begin{equation}
		\lim_{N,N_{\text{modes}}\rightarrow\infty}\mathbb{E}_{\kappa \sim \mathcal{D}}\left[ \| \psi-\hat{\psi}\|_{L^2(\Omega)} \right]=0
	\end{equation}
    \solidqed

\section{Numerical Experiments}
\label{results}

\subsection{Parameter Setting}
Our numerical experiments were conducted in an environment equipped with an NVIDIA Tesla V100 PCIE GPU, utilizing the PyTorch deep learning framework, and an Intel(R) Xeon(R) Silver 4210 CPU for numerical computation.

The key hyperparameters used in model training are as follows:
\begin{enumerate}
	\item \textbf{Learning rate}: $1 \times 10^{-4}$
	
	\item \textbf{$L^2$ regularization coefficient}: $1 \times 10^{-4}$
	
	\item \textbf{Number of training epochs}: 100
	
	\item \textbf{Fourier mode coefficients}: 12 along the X-axis, 12 along the Y-axis
\end{enumerate}

These parameter choices were carefully selected to ensure stable training and effective model performance.

\subsection{Ablation Study}
Ablation study is an efficient method to investigate the superiority of the model and to evaluate the contribution of each component. In our study, to assess the effectiveness of the preconditioner and the advantage of the multiscale neural operator compared to a single-scale learner, we conduct an ablation study by removing certain parts of the architecture to evaluate their impact on the learning of basis functions. To ensure experimental rigor, we adhere to the single-variable principle, where only one component (either the preconditioner, coarse-grid path, or fine-grid path) is removed at one time, while keeping all other components unchanged. The model is evaluated using the same metrics as the complete model, specifically MSE, MAE, and $\text{R}^2$.

\textbf{Ablation 1: Removing the preconditioner}. The preconditioner is a crucial component of our proposed method. Removing this part causes the permeability field to be directly fed into the multiscale operator for feature learning. To maintain structural consistency, we use a $1 \times 1$ filter to process the data, ensuring that the output channel count matches that of the original preconditioner. All other structures and parameters, such as the number of training epochs and optimizer settings, remain unchanged.

\textbf{Ablation 2 and 3: Removing a Single-Scale Neural Operator}. The multiscale operator in our architecture learns features at different scales. To investigate the importance of different scales within the entire operator, we individually remove the coarse-grid path and fine-grid path. The primary impact of this modification on the structural design is that the output is directly fed into the fully connected layer without requiring convolutional fusion. For consistency, we continue using L2 regularization in the fully connected layer to simulate Ridge Regression. All other parameters and structures remain unchanged. 

The three models resulting from the different ablation methods are denoted Ablation1 (Preconditioner), Ablation2 (coarse) and Ablation3 (Fine), respectively. The specific experimental results will be presented and analyzed in the Results section.

\begin{table}[h]
	\centering
	\caption{Model Performance Metrics: MSE, MAE and $\text{R}^2$, with standard deviation.}
	\renewcommand{\arraystretch}{1.2} 
	\noindent\hspace*{-0.5cm} 
	\makebox[\textwidth][c]{ 
		\begin{tabular}{ccccc}
			\hline
			\textbf{Subset} & \textbf{Model} & \textbf{MSE} & \textbf{MAE} & \textbf{R²} \\
			\hline
			\multirow{4}{*}{Training} & Full & $0.0080 \pm 9.3100\times 10^{-10}$ & $0.0512 \pm 7.4500\times 10^{-9}$ & $0.9566 \pm 5.4600\times 10^{-8}$ \\
			& Ablation1 & $0.0341 \pm 5.5255\times 10^{-9}$ & $0.1129 \pm 1.6990\times 10^{-8}$ & $0.8436 \pm 5.6231\times 10^{-8}$ \\
			& Ablation2 & $0.0080 \pm 1.5584\times 10^{-9}$ & $0.0502 \pm 9.3504\times 10^{-9}$ & $0.9554 \pm 1.7881\times 10^{-8}$ \\
			& Ablation3 & $0.0320 \pm 3.7253\times 10^{-9}$ & $0.1052 \pm 8.1617\times 10^{-9}$ & $0.8400 \pm 7.0777\times 10^{-8}$ \\
			\hline
			\multirow{4}{*}{Validation} & Full & $0.0155 \pm 0.0000$ & $0.0708 \pm 0.0000$ & $0.9310 \pm 0.0000$ \\
			& Ablation1 & $0.0554 \pm 0.0000$ & $0.1474 \pm 0.0000$ & $0.7774 \pm 0.0000$ \\
			& Ablation2 & $0.0156 \pm 9.3132\times 10^{-10}$ & $0.0707 \pm 7.4506\times 10^{-9}$ & $0.9292 \pm 0.0000$ \\
			& Ablation3 & $0.0686 \pm 0.0000$ & $0.1636 \pm 0.0000$ & $0.7098 \pm 0.0000$ \\
			\hline
			\multirow{4}{*}{Testing} & Full & $0.0036 \pm 0.0000$ & $0.0375 \pm 0.0000$ & $0.9716 \pm 0.0000$ \\
			& Ablation1 & $0.0206 \pm 1.8626\times 10^{-9}$ & $0.0887 \pm 0.0000$ & $0.8827 \pm 0.0000$ \\
			& Ablation2 & $0.0035 \pm 2.3283\times 10^{-10}$ & $0.0357 \pm 0.0000$ & $0.9709 \pm 0.0000$ \\
			& Ablation3 & $0.0140 \pm 9.3132\times 10^{-10}$ & $0.0697 \pm 0.0000$ & $0.9041 \pm 0.0000$ \\
			\hline
			\multirow{4}{*}{Total} & Full & $0.0086 \pm 9.3132\times 10^{-10}$ & $0.0524 \pm 0.0000$ & $0.9537 \pm 0.0000$ \\
			& Ablation1 & $0.0357 \pm 0.0000$ & $0.1150 \pm 7.4506\times 10^{-9}$ & $0.8368 \pm 0.0000$ \\
			& Ablation2 & $0.0086 \pm 0.0000$ & $0.0514 \pm 3.7253\times 10^{-9}$ & $0.9524 \pm 0.0000$ \\
			& Ablation3 & $0.0357 \pm 0.0000$ & $0.1098 \pm 7.4506\times 10^{-9}$ & $0.8227 \pm 0.0000$ \\
			\hline
		\end{tabular}
	}
	\label{tab_criteria}
\end{table}

\begin{figure}[h]
	\centering
	\includegraphics[width=0.7\linewidth]{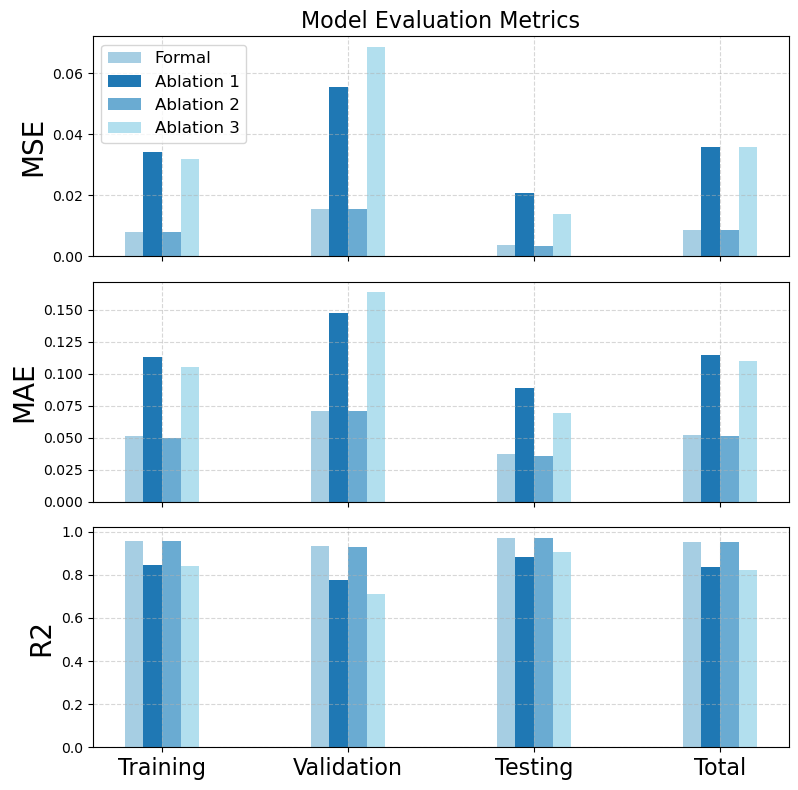}
	\caption{Histogram plot of the model performance, including training set, validation set, testing set and total dataset.}
	\label{fig_criteria}
\end{figure}
Table.\ref{tab_criteria} and Figure.\ref{fig_criteria} present the evaluation metrics of the model and ablation studies in two different ways.

(\textbf{Impact of Preconditioner}) Ablation 1 represents the model without the preconditioner. As shown in Table.\ref{tab_criteria}, the evaluation metrics indicate that the model's MSE, MAE and $\text{R}^2$ on the testing set are 0.0206, 0.0887 and 0.8827, respectively, compared to 0.0036, 0.0602 and 0.9716 for the full model. This comparison suggests that the preconditioner significantly enhances the solver's ability to predict the multiscale basis functions. This architecture setup effectively improves the learning of features in the permeability fields.

(\textbf{Impact of Multiscale Pathways}) Ablation 2 and 3 represent ablation studies in which the fine and coarse grid systems are removed, respectively. The metrics for Ablation 2 on the testing set are 0.0035, 0.0357 and 0.9709 for MSE, MAE and $\text{R}^2$, while the corresponding metrics for Ablation 3 are 0.0357, 0.1098 and 0.8227. Recall the fact that our basis functions are defined on the coarse-grid system. The exclusion of feature learning on the coarse grid has a significant impact on the overall model performance. Although the removal of fine-grid path did not result in a statistically significant decline in model performance, the fine-grid system still contains relevant information, which can be shown in the value of variances.

\begin{figure}[!h]
	\centering
	\includegraphics[width=1.0\linewidth]{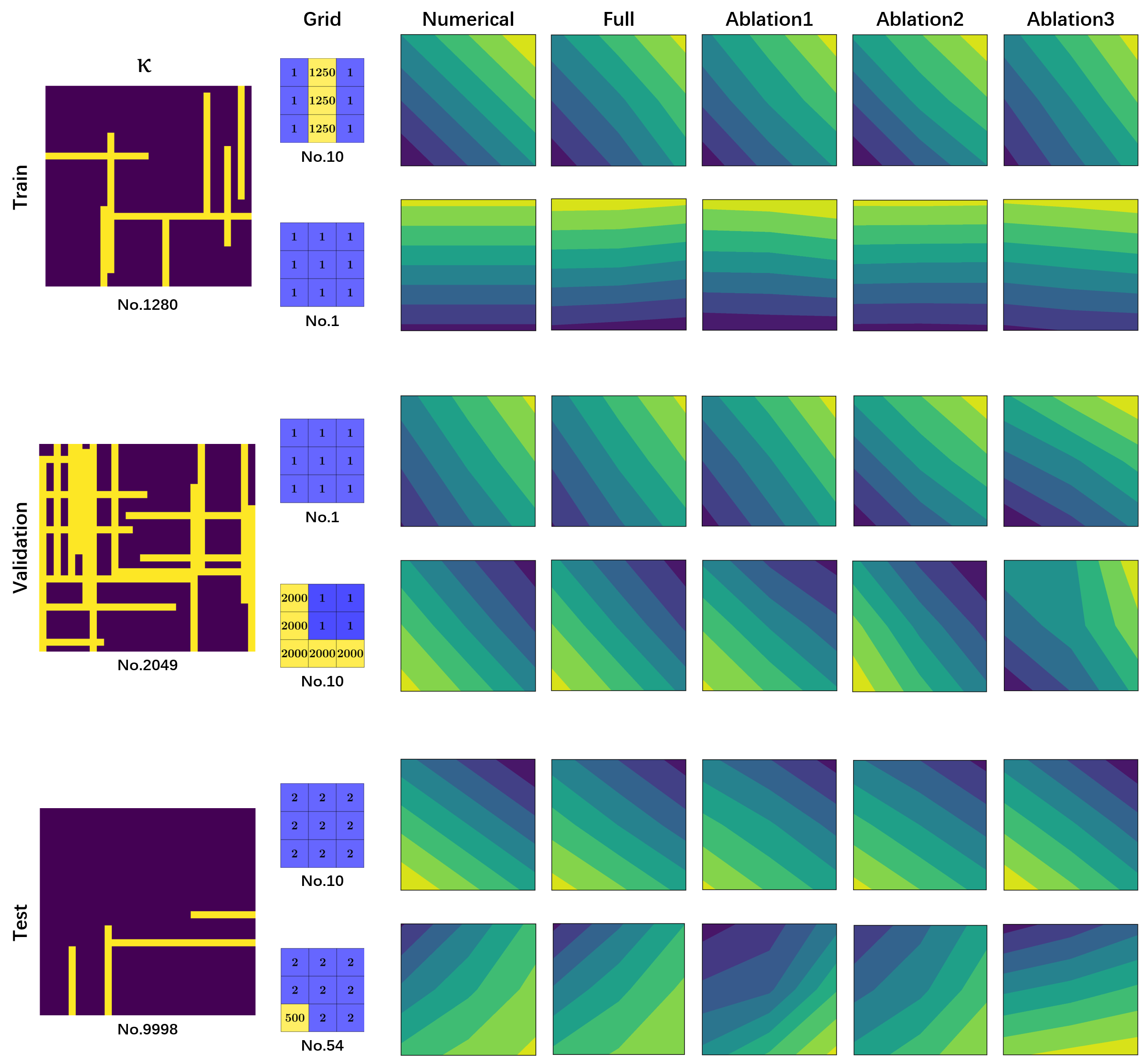}
	\caption{Contour Plot of single coarse grid predicted by different models. From up to Bottom: Data sampled from training, validation and testing set, respectively. From left to right: 
		(\textbf{Column 1}) High-contrast permeability field $\kappa$. The caption "No.n" means that the data is the n-th sample of corresponding dataset. 
		(\textbf{Column 2}) Selected example coarse grid. The caption "No.m" means the m-th coarse grid of the permeability field. The yellow in the grid represents the fracture and the blue represents the matrix, where the numbers are the corresponding permeability (millidarcy). (\textbf{Column 3}) Numerical results of mixed GMsFEM. (\textbf{Column 4}) Deep learning results of the full model. 
		(\textbf{Column 5}) Results of ablation 1 model, which removes the preconditioner block. 
		(\textbf{Column 6}) Results of ablation 2 model, which removes the fine-path network. 
		(\textbf{Column 7}) Results of ablation 3 model, which removes the coarse-path network.}
	\label{fig_results}
\end{figure}

\begin{figure}[h]
	\centering
	\includegraphics[width=1.0\linewidth]{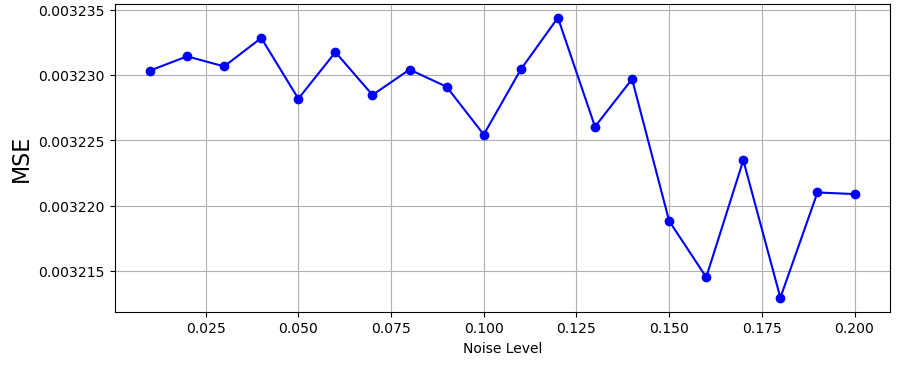}
	\caption{Stability assessment for the model. Adding noise to the one batch of the testing dataset and observe the change of MSE in the evalution metrics of the observation model.}
	\label{fig_stability_mse}
\end{figure}

The prediction results are shown in Figure.\ref{fig_results}, which demonstrates the contour plots of both fractured and non-fractured coarse grids. In these grids, blue and yellow represent matrix and fracture, respectively. Column 4-7 show the predictions generated by different models. It can be seen that there exists significantly different plots in Ablation 1 and 3 (column 5 and 7), especially for the grid with fractures. This figure, together with performance metrics, can show the effectiveness of the preconditioner and multiscale network. For homogenized permeability fields, this architecture demonstrates outstanding performance, which assists a single CNN model in extracting feature efficiently.

\subsection{Stability Analysis}
\begin{figure}[h]
	\centering
	\includegraphics[width=1.0\linewidth]{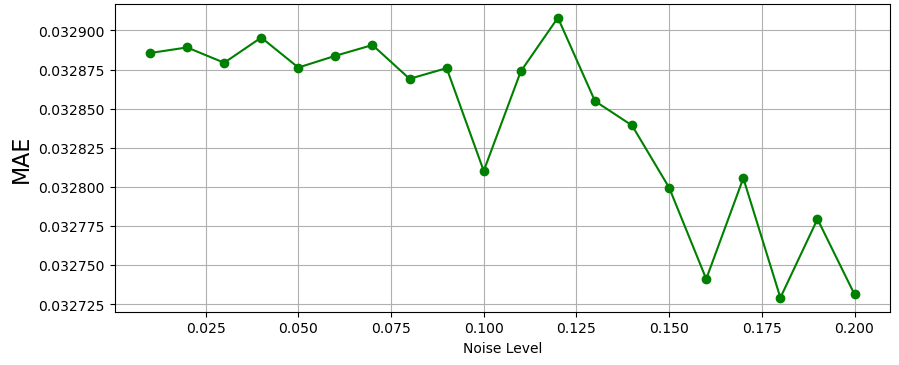}
	\caption{Stability assessment for the model. Adding noise to the one batch of the testing dataset and observe the change of MAE in the evalution metrics of the observation model.}
	\label{fig_stability_mae}
\end{figure}

To evaluate the stability of the model, we analyzed the variations in the evaluation metrics MSE, MAE, and $\text{R}^2$ under the influence of 20 different levels of noise (the same distribution as how we generate the permeability fields) ranging from 0.01 to 0.2. 
As shown in Figure.\ref{fig_stability_mse}, the MSE exhibits a fluctuating yet gradually decreasing trend with increasing noise levels, with changes consistently remaining within the $10^{-4}$ range. 
Similarly, Figure.\ref{fig_stability_mae} demonstrates that MAE experiences a fluctuating decline under different noise disturbances, also confined within $10^{-4}$ range, indicating that the model maintains high accuracy despite varying degrees of noise interference. 

\begin{figure}[h]
	\centering
	\includegraphics[width=1.0\linewidth]{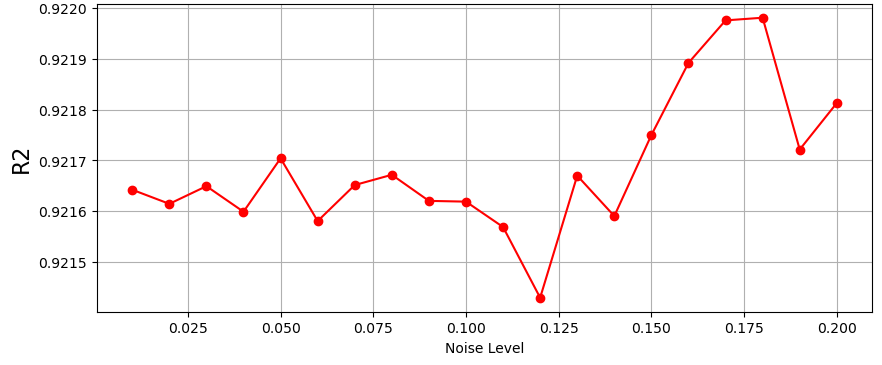}
	\caption{Stability assessment for the model. Adding noise to the one batch of the testing dataset and observe the change of $\text{R}^2$ in the evalution metrics of the observation model.}
	\label{fig_stability_r2}
\end{figure}

In contrast, $\text{R}^2$, as depicted in Figure.\ref{fig_stability_r2}, displays a fluctuating pattern and eventually rises as noise levels increase, with its overall change remaining within $10^{-4}$ magnitude. 
These findings indicate that the model exhibites stability under Gaussian noise interference, demonstrating an ability to resist noise and maintain good ability in predictions.

\begin{figure}[h]
	\centering
	\includegraphics[width=1.0\linewidth]{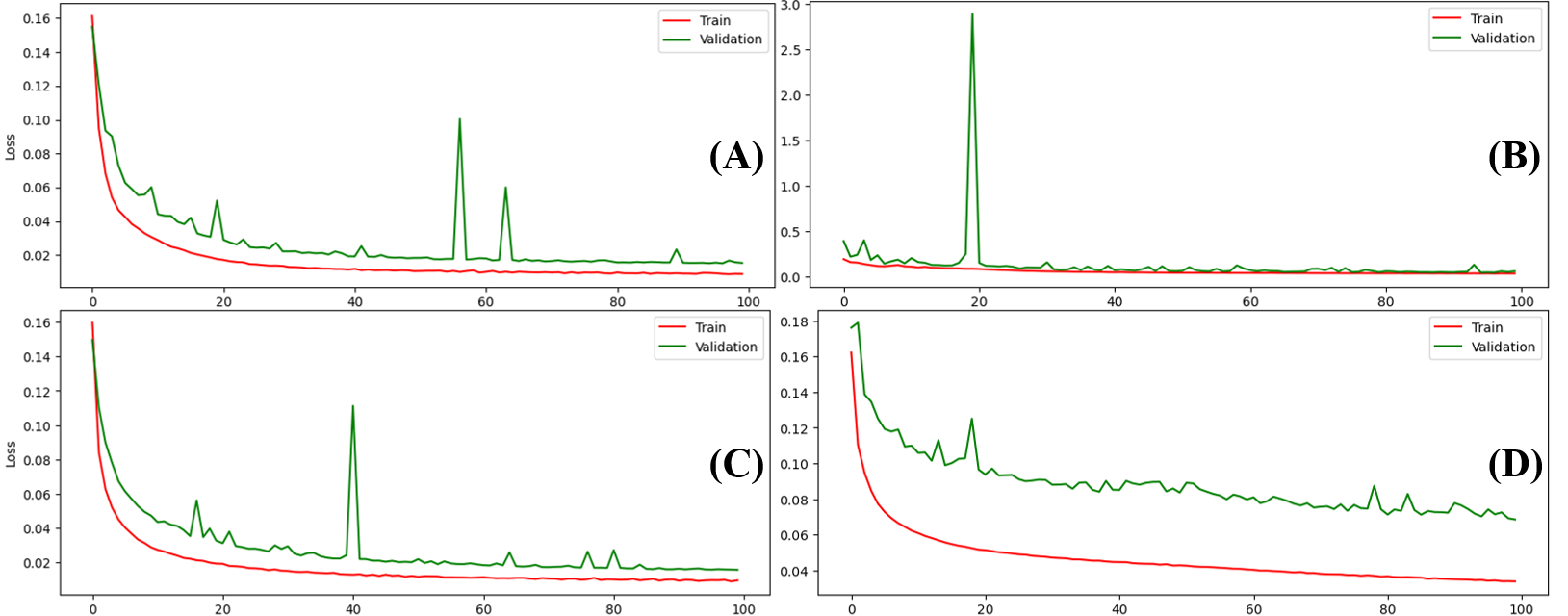}
	\caption{Learning curves of our full model and the ablation studies. (A)-Full model. (B)-Ablation 1, whose preconditioner block is removed. (B)-Ablation 2, which has only coarse-path network. (C)-Ablation 3, which has only the fine-pathway natwork. Red and green refer to training and validation loss, respectively.}
	\label{fig_learning_curve}
\end{figure}

As one of the components to assess the convergence of the training process, learning curve is a useful tool to observe the change of training and validation loss. 
As shown in Figure.\ref{fig_learning_curve}, it can partially show whether there exists overfitting in the training. 
In Figure.\ref{fig_learning_curve}-(A), our proposed model shows a consistent trend between training and validation losses, maintaining stability throughout the training process. 
The validation loss gradually decreases to a relatively low level, indicating that the model exhibits a good fitting ability on both training and validation sets, thereby demonstrating excellent generalization capability. 

In contrast, the Ablation 1 model, shown as Figure.\ref{fig_learning_curve}-(B), exihibits a significant outlier in the validation loss, reaching as high as near 3.0, indicating severe overfitting or instability on the validation set, thus reflecting poor generalization capability. 
The Ablation 2 model in Figure.\ref{fig_learning_curve}-(C) and Ablation 3 model in Figure.\ref{fig_learning_curve}-(D) both show some level of fluctuation in the validation loss. 
Ablation 2 shows similar curve with the full model, while Ablation 3 model have significant difference between the two loss arrays, indicating that the model without coarse-path network can not extract sufficient features from the data.

\subsection{Comparison of Prediction Time Consumption}
Since the training process of neural networks requires a certain amount of time, we compare the computational efficiency of the generated neural network model with that of the mixed GMsFEM method in computing multiscale basis functions. To ensure a fair comparison, we pre-generate the corresponding permeability field data, avoiding additional time consumption from random field generation. Specifically, we compare the time required for a single computation (or prediction) as well as the total time for multiple repeated computations, with the corresponding results presented in Table.\ref{tab_time}.

\begin{table}[h]
	\centering
	\caption{Comparison of Time Consumption (s).}
	\setlength{\tabcolsep}{6pt} 
	\begin{tabular}{p{3cm} p{2.5cm} p{2.5cm}}
		\toprule
		\textbf{Method} & \textbf{1 time} & \textbf{100 times} \\
		\midrule
		FP-HMsNet & 0.022 & 3.678  \\
		mixed GMsFEM &  1.388 & 67.082 \\
		\bottomrule
	\end{tabular}
	\label{tab_time}
\end{table}

These findings demonstrate that our proposed model maintains stable performance across different datasets, exhibiting lower validation losses and no significant fluctuations. This confirms its superior stability and convergence, even in the presence of noise interference and increased model complexity. Moreover, our results indicate that deep learning models achieve significantly higher computational efficiency, particularly when performing repeated computations or prediction tasks, compared to traditional numerical algorithms. This highlights the clear advantage of our method in terms of efficiency, making it a more practical and scalable solution for complex multiscale problems.

\section{Discussion}
\label{discussion}

This study proposes a more accurate and efficient network architecture for predicting multiscale basis functions of subsurface fluid flow in porous media. 
By introducing a lightweighted preconditioner to the multiscsale neural network, our model demonstrates superior performance acoross evaluations of accuracy, convergence and stability, aligning well with our expectations.

The performance enhancement can be attributed to the preconditioner's ability to map input data into the frequency domain, thereby extracting discriminative spectral features critical for deep learning frameworks. Theoretical analysis and experimental validation demonstrate that preprocessing data with such frequency domain transformations before feeding them into a CNN significantly improves the efficiency of feature learning. Notably, the proposed multiscale CNN architecture employs a hierarchical filter bank to extract cross-resolution features. Crucially, this framework integrates multiscale features with basis functions defined on coarse-grid systems, a design choice validated by ablation studies to be pivotal for model performance. The multiscale hierarchical structure further ensures prediction stability by constructing a feature pyramid with stable representational capacity.

From a computational perspective, the frequency-domain transformation induced by the preconditioner reduces convolutional operations to linear transformations in the spectral space. This reformulation lowers the time complexity from $\mathcal{O}(LK^2N^2) + \mathcal{O}(N^2F)$ in the spatial domain to $\mathcal{O}(N^2\log N)+\mathcal{O}(N^2F)$, achieving substantial computational savings. Such efficiency gains highlight the method's suitability for memory-constrained environments and latency-sensitive applications. Empirical results corroborate the algorithm's stability and practical utility under resource-limited conditions.

Compared to previous studies, our research aligns with established findings that deep learning techniques can be effectively applied for predicting multiscale basis functions with high accuracy and low information losses. 
Importantly, the evaluation on the testing set shows that our model achieves an MSE of 0.0036 and $\text{R}^2$ of 0.9716, while the corresponding results from traditional CNN model of the previous research were 0.0446 and 0.8083, respectively \cite{choubineh2022innovative}. These outcomes clearly indicate that our model surpasses existing models in terms of accuracy. The introduction of preconditioner and multiscale parallel learning strategy allow our model to capture more features and effectively predict the basis functions. 

While the current findings demonstrate promising potential, several research extensions warrant further investigation. 
Primarily, extending the model to characterize three-dimensional anisotropic permeability fields could significantly enhance its applicability in realistic subsurface environments (e.g., heterogeneous reservoirs). 
Furthermore, developing a multi-physics coupling framework to resolve cross-scale interactions between geological formations (e.g., fault-fracture networks) would advance mechanistic understanding. Rigorous validation using real-world reservoir data (e.g., well logs and seismic inversion datasets) remains essential to verify engineering practicality. These advancements may catalyze paradigm shifts in subsurface flow modeling and establish novel methodologies for intelligent mineral exploration in deep geological settings.

This study acknowledges its limitations: the training dataset is primarily focused on fractured porous media, and the effectiveness of this strategy when applied to other types of porous media remains to be discussed. Nonetheless, theoretical analysis reveals the model's capability to adaptively construct multi-resolution basis functions from unstructured permeability fields. This core algorithm offers an interpretable computational tool for probabilistic inversion of subsurface mineral distributions, such as identifying rare-earth element enrichment zones.

\section{Conclusion and Future Work}
\label{conclusion}

This study proposes FP-HMsNet, a framework with a preconditioner-learner architecture, integrating Fourier neural operators with multiscale networks for predicting multiscale basis functions in fractured porous media. 
By synergistically capturing global and local spatial features, the model demonstrates exceptional capability in resolving flow patterns within fracture-matrix systems. 
The hierarchical architecture, enhanced by the preconditioner's spectral decoupling mechanism, establishes a paradigm for high-fidelity fracture network modeling, as rigorously validated through ablation studies.

While the current work focuses on 2D flow simulations in fractured media, future research will extend to 3D, large-scale heterogeneous systems (e.g., carbonate pore networks and shale nanoporous structures) to comprehensively evaluate geological adaptability. 
Additionally, adaptive multiresolution architectures will be developed for coupled flow-transport processes across scales. 
These advancements aim to bridge theoretical modeling with engineering applications, particularly in unconventional hydrocarbon recovery and subsurface environmental monitoring.

\section*{Acknowledgment}
The authors wish to thank the anonymous referees for their thoughtful comments, which helped in the improvement of the presentation.

\section*{CRediT Authors Contributions Statement}
\textbf{Jie Chen}: Conceptualization, Data curation, Formal analysis, Methodology, Software, Supervision, Validation, Writing-review \& editing. \textbf{Peiqi Li}: Conceptualization, Methodology, Project administration, Visualization, Writing-original draft \& review \& editing. \textbf{Zhengkang He}: Data curation, Proof reading \textbf{Hands Simon}: Supervision, Proof reading.

\section*{Declaration of Competing Interests}
The authors confirm that they have no conflict of interest in relation to the content of this study. 

\section*{Data and Code Availability}
Due to proprietary considerations, both the dataset and the analysis code used in this study are not publicly available. Researchers interested in accessing the data or the code may contact the corresponding author for more information.

\bibliographystyle{elsarticle-harv}
\bibliography{references.bib}

\end{document}